\documentstyle[12pt]{article}
\newcommand{\beq}{\begin{equation}}
\newcommand{\eeq}{\end{equation}}
\begin{document}

\thispagestyle{empty}

\begin{flushright}
{\parbox{3.5cm}{
UAB-FT-365

May, 1995

hep-ph/9505350
}}
\end{flushright}

\vspace{3cm}
\begin{center}
\begin{large}
\begin{bf}
MATCHING THE LOW AND HIGH ENERGY DETERMINATIONS OF $\alpha_s(M_Z)$ IN
THE MSSM\\
\end{bf}
\end{large}
\vspace{1cm}
David GARCIA, Joan SOL\`A\\
\vspace{0.25cm}

Grup de F\'{\i}sica Te\`orica\\
and\\
Institut de F\'\i sica d'Altes Energies\\
\vspace{0.25cm}
Universitat Aut\`onoma de Barcelona\\
08193 Bellaterra (Barcelona), Catalonia, Spain\\
\end{center}
\vspace{0.3cm}
\hyphenation{super-symme-tric}
\hyphenation{com-pe-ti-ti-ve}
\begin{center}
{\bf ABSTRACT}
\end{center}
\begin{quotation}
\noindent
Recent calculations of supersymmetric corrections to the
conflicting ratios $R_b$ and $R_c$ have shown that an alleged discrepancy
between the SM predictions of these observables
and the corresponding experimental values
can be cured in the MSSM
within a certain region of the parameter space.
Here we show that, in this very same region,
also a well-known discrepancy between the low and high energy
determinations of $\alpha_s(M_Z)$ can be disposed of. Specifically, we find
that the lineshape determination of the strong coupling constant, which
in the SM points towards the large central value
$\alpha_s(M_Z) \stackrel{\scriptstyle >}{{ }_{\sim}} 0.125$, can
be matched up with the value  suggested by the
wealth of low-energy data, namely $\alpha_s(M_Z) \simeq 0.11$, which is
smaller
and more in line with the traditional QCD expectations at low energy.
Our approach differs from
previous analyses in that we argue that the desired matching could
originate
to a large extent from a purely electroweak supersymmetric quantum effect.
\end{quotation}

\baselineskip=6.5mm  

\newpage

The world average of the various determinations of the strong coupling
constant at the scale of the $Z$-boson mass within the Standard
Model (SM) reads as follows
\,\cite{Bethke}:
\beq
\alpha_s(M_Z)=0.118\pm 0.007\,.
\label{eq:alphaglobal}
\eeq
However, one should not forget that this number follows from a gross
averaging
of several groups of data of very different nature whose respective central
values
show up statistically significant differences.
Hence eq.(\ref{eq:alphaglobal}) is suspicious of hidding a potentially
important problem within the SM. Even more: it could prompt us about
physics beyond the SM,
as has already been pointed out by several authors
\,\cite{ErlerLang,Shifman}. Indeed, one finds that the high-energy data,
i.e. data taken directly at $q^2=M_Z^2$ from lineshape and
event shape analyses at LEP and SLD, turn out to cluster around a ``large''
value of $\alpha_s(M_Z)$. For example, the lineshape value
is\,\cite{Martinez1}
\beq
\alpha_s(M_Z)=0.126\pm 0.007.
\label{eq:alphahigh}
\eeq
In contrast, the low-energy data obtained from a great variety of
sources (such as deep inelastic scattering,
lepton scattering and quarkonium spectra)
concentrates\footnote{ Upon due account, of course, of the
renormalization group running from the low-energy scale
$q^2<<M_Z^2$ where the data have been taken
up to $q^2=M_Z^2$.} around a central value which is about $12\%$ smaller.
For instance, the deep inelastic scattering result reads\,\cite{Bethke}
\beq
\alpha_s(M_Z)=0.112\pm 0.005\,.
\label{eq:alphalow}
\eeq
We infer that the low-energy determination of $\alpha_s(M_Z)$
lies three standard deviations below the high-energy determination.
Although the latter is consistent with the value obtained
from  the $\tau$-decay fraction measurement (a low-energy result),
here what is in dispute is the theoretical
method of calculation\,\cite{tau} and the treatment of errors
\cite{Shifman,Altarelli}.

The clash between the low and high energy determinations
of $\alpha_s(M_Z)$ is a persisting result in the literature
\,\cite{Bethke}
and it has become a controversial and disturbing problem
(a sort of  ``$\alpha_s(M_Z)$ crisis'') within the
context of the standard model picture of the strong interactions.
Recently, Shifman has drawn much attention
on this issue\,\cite{Shifman}.
He strongly advocates in favour of $\alpha_s(M_Z)\simeq 0.11$--the
deep inelastic result, eq.(\ref{eq:alphalow}), being a case in point
\footnote{Lattice calculations also agree with the low-energy
value (\ref{eq:alphalow}) and
they produce $\alpha_s(M_Z)=0.115$ or less \,\cite{Kronfeld}.}--
and argues that the large result (\ref{eq:alphahigh}) is incompatible with
crucial features of QCD; for example --following Shifman's contention--,
a strong coupling constant $\alpha_s(M_Z)>0.11$
would entail $\Lambda_{QCD}>200\,MeV$,
where $\Lambda_{QCD}\equiv \Lambda_{\overline{MS}}^{(4)}$
--suitable for deep inelastic scattering.
In particular, for $\alpha_s(M_Z)\simeq 0.125$ one has to endure a
corresponding value of
$\Lambda_{QCD} \simeq 500\,MeV$
which is considered far too large to explain the evolution of moments
of the structure functions in deep inelastic scattering and
the success of the operator product expansion in a wide range of QCD
problems.
On these grounds, solid and convincing as they may be,
Shifman suggests that the large lineshape result (\ref{eq:alphahigh})
claims for new physics, in particular for unconventional strong
interations,
such as e.g. supersymmetric interactions mediated  by gluinos and squarks,
which have not been taken into account in the theoretical calculation
of the hadronic $Z$-width.  Upon inclusion
of these contributions, the high-energy value (\ref{eq:alphahigh}) should
descend to the preferred low-energy value (\ref{eq:alphalow}).
On the basis of existing calculations
of the SUSY-QCD corrections to the $Z$-width\,\cite{Djouadi} one
can sustain that this could be the case provided that gluinos have a mass
of order of a few GeV and squarks are in the ballpark of $100\,GeV$.
We remark that for ${\cal O}(100)\,GeV$ gluinos, non-negligible corrections
are also possible in certain regions of parameter space, but they
invariably come out with the ``wrong'' sign, i.e. they would induce a
lineshape
value of $\alpha_s(M_Z)$ even larger than that of eq.(\ref{eq:alphahigh})
(see later on).

As for the light gluino scenario, it
was already advocated long ago
in the literature\,\cite{Kowalski}, and also more recently by
several authors\,\cite{Clavelli}, though in all these cases
from a different point of view:
namely, as a means to slow down the renormalization group
running of $\alpha_s(q^2)$ from the low-energy scale up to $q^2=M_Z^2$
due to extra, negative, contributions to the strong coupling
$\beta$-function.
Unfortunately, the extremely narrow slit for the existence of these
very light gluinos of ${\cal O}(1)\,GeV$ is nearly--if not completely--
closed experimentally.

It should be pointed out that
alternative (non-SUSY) scenarios have also been proposed to cure the
``$\alpha_s(M_Z)$ crisis'', such as e.g. the
extended technicolor approaches
of Ref.\cite{ETC}. However, it remains to see whether they are, too,
consistent with other observables like $R_b$ and $R_c$ (see below).
Moreover, in contrast to the MSSM,
technicolour models are not in very good shape to fit the plethora of
available electroweak precision data at the moment\,\cite{EFL}.

On the other hand, from the point of view of grand desert supersymmetric
Grand Unified Theories (SUSY-GUT's),
the typical range of large values (\ref{eq:alphahigh})
also obtains in all renormalization group
analyses based on canonical assumptions\,\cite{LangPol1}
\footnote{It is worth remembering that in non-SUSY-GUT scenarios the
corresponding prediction
is unacceptably too small: $\alpha_s(M_Z)=0.073\pm
0.002$\,\cite{LangPol2}.}.
However, as Ref.\cite{Roszkowski} shows, one can try to reconcile
SUSY-GUT's with the low-energy value (\ref{eq:alphalow}) at the price of
relaxing several common place assumptions on gaugino mass relations
and in general by adopting a more phenomenological attitude towards
the structure of the Minimal Supersymmetric Standard Model
(MSSM)\,\cite{MSSM}.
In a sense this is also the philosophy adopted in the rather different
context of
Refs.\cite{GJS1}-\cite{GJS3}, where we systematically searched for the
pattern of
supersymmetrical particle masses preferred by phenomenology in order
to solve an apparent
discrepancy between theory and experiment in the ratio
\beq
R_b={\Gamma(Z\rightarrow b\bar{b})
\over \Gamma(Z\rightarrow {\rm hadrons})}\,,
\label{eq:Rb}
\eeq
where $\Gamma(Z\rightarrow{\rm hadrons})\equiv\sum_{q=u,d,c,s,b}
\Gamma(Z\rightarrow q\bar{q})$ stands for the partial $Z$-width into
hadrons resulting from a primarily produced quark-antiquark pair.
Within the context of the SM, and for the present range of values of the
top quark
mass\,\cite{topmass}, $R_b$ is predicted to be {\it below} the
experimental result\footnote{ See e.g. Ref.\cite{Martinez1} for a complete
report on the status of the $Z$-physics observables.}.
In contrast, as
it follows from our analysis\,\cite{GJS2} as well as from
previous analyses in the literature\,\cite{Rb},
the theoretical prediction of $R_b$ in the MSSM could result in a net
increase
in the value of this observable.
As a drawback, however,
the theoretical improvement of $R_b$ can only be achieved within the
framework of a general MSSM with a minimum number of assumptions on its
spectrum of sparticle masses. In other words, the supersymmetric prediction
of $R_b$ does not substantially improve in the context of
``canonical'' SUSY-GUT's, such as e.g. in the so-called
constrained MSSM\,\cite{CMSSM} or in the simplest (and next to
simplest)  supergravity
(SUGRA) scenarios\,\cite{RbSUGRA}. Still, as already
mentioned, one can try to amend this situation in non-minimal
(e.g. string-like) SUSY-GUT's
\,\cite{Roszkowski}; after all, the minimal SUGRA models,
although provide an economic and elegant framework, they might be based
on oversimplified assumptions about physics at the unification scale, and
they may ultimately prove to be incorrect.

Further studies of the authors\,\cite{GS} in the framework of the MSSM
have shown that one can improve the theoretical
prediction, not only of the ratio $R_b$, but also of the companion ratio
\beq
R_c={\Gamma(Z\rightarrow c\bar{c})
\over \Gamma(Z\rightarrow {\rm hadrons})}\,.
\label{eq:Rc}
\eeq
In contradistinction to $R_b$,
this observable is predicted in the SM to be {\it above} the experimental
measurement\,\cite{Martinez1}.
Quite intriguingly, it turns out that upon taking into account the
supersymmetric electroweak quantum effects predicted by the general MSSM
in a particular region of parameter space, the theoretical prediction can
be substantially ameliorated\,\cite{GS}. And what is more, one can
{\it simultaneously } alleviate the  $R_b$ and $R_c$ ``crises'' in the
MSSM.
For this to be so it suffices that the following
conditions are fulfilled:

i) $\tan\beta$ should be large enough, namely
$\tan\beta\stackrel{\scriptstyle >}{{ }_{\sim}}{\cal O}(m_t/m_b)$;

ii) There should exist a light
supersymmetric CP-odd (``pseudoscalar'') Higgs, $A^0$, of ${\cal
O}(50)\,GeV$
($m_{A^0}<70\,GeV$, to avoid negative effects from charged
higgses\,\cite{GJS2});

iii) For a comfortable solution, there should also exist
 a light chargino, $\Psi^{\pm}_1$,
in the $50-60\,GeV$ ballpark\,\cite{GJS2,GS}\,,

We point out that similar results can also
be obtained for very large $m_{A^0}$, i.e. for effectively
decoupled non-standard higgses,  and very low values
of $\tan\beta$, typically $\tan\beta<0.7$, which insure a big contribution
from the sparticles alone. In this case condition iii) above has to be
supplemented with the requirement of
a light stop whose mass is of the same order of magnitude as the
chargino mass.
However, the range $\tan\beta<1$ is currently not supported by model
building
and we shall not consider this possibility as our favourite Ansatz.

In spite of not being strictly necessary for the $Z$ observables under
study,
one might also like to have
a light stop, $\tilde{t}_1$, even at high $\tan\beta$. This could be
necessary\,\cite{Garisto}
to make the MSSM compatible with the CLEO bound
on $B(b\rightarrow s\,\gamma)$\,\cite{CLEO}.
In this respect it should be recalled that light stops have been
invoked in previous analyses of $R_b$ in the MSSM which emphasized
the region of moderate and small values of
$\tan\beta$\,\cite{Rb}. However, in a regime of large $\tan\beta$, a light
stop is not essential to account for the ratios $R_b$ and $R_c$ themselves,
as explicitly shown in Refs.\cite{GJS2,GS}.
Moreover, postulating a light stop on the
sole basis of demanding consistency with the CLEO bound may not be too
compelling; after all, a satisfactory treatment of conventional QCD
corrections
to $B(b\rightarrow s\,\gamma)$, which are known to be very large,
is still lacking\,\cite{QCDCLEO}. Therefore, we
should better be openminded at  this point.

In the light of the above considerations, it would be desirable to
explore whether the electroweak supersymmetric corrections may account
at the same time for the disturbing ``$\alpha_s(M_Z)$ crisis'' and
to  assess whether the purported solution space for the
``$R_b-R_c$ crisis'' does overlap with that of the ``$\alpha_s(M_Z)$
crisis'',
in which case the relevance of the supersymmetric solution would be
significantly augmentated.

In Figs.1-4 we deliver the final result of our analysis, which is based on
the
detailed calculations of all oblique and non-oblique one-loop
supersymmetric
electroweak quantum effects on the partial $Z$-widths into
fermion-antifermion
pairs in the MSSM, as presented in Refs.\cite{GJS1,GJS2} whose notation and
definitions we shall adopt henceforth.
In the case under consideration
it will suffice to consider the decay modes involved in
the ratio of the hadronic and electronic partial
 $Z$ widths:
\beq
R\equiv{\Gamma_h\over\Gamma_e}\equiv
{\Gamma(Z\rightarrow {\rm hadrons})\over \Gamma(Z\rightarrow e^+e^-)}\,.
\label{eq:R}
\eeq
This observable
is the relevant quantity in our analysis, for it is directly
involved in the hadronic peak cross-section
\beq
\sigma_h^0\equiv
\sigma (e^+e^-\rightarrow{\rm hadrons})|_{s=M_Z^2}={12\pi\over M_Z^2}
\left({\Gamma_e\over \Gamma_Z}\right)^2 R
\label{eq:peak}
\eeq
from which the lineshape determination of $\alpha_s(M_Z)$
ensues.
In the SM one computes
$\Gamma_h^{SM}=\Gamma_h^{SM}(\alpha_s(M_Z))$ as a function of
$\alpha_s(M_Z)$,
and by comparing it with the experimentally measured lineshape value of $R$
(notice that $R=\Gamma_h/\Gamma_e=\sigma_h^0/\sigma_e^0$)
one derives the result (\ref{eq:alphahigh}), where the error reflects
uncertainties on the electroweak and QCD parts of the theoretical
prediction,
as well as on the lack of knowledge of the Higgs mass (assumed in the range
$60-1000\,GeV$, with central value $m_H=300\,GeV$) and of
the top quark mass whose determination is still rather
poor\,\cite{topmass}.

In the context of the MSSM it is convenient to cast each partial
width $\Gamma(Z\rightarrow f\bar{f})$ split up as follows:
\beq
\Gamma_Z^{MSSM}=\Gamma_Z^{RSM}+\delta\Gamma_Z^{MSSM}\,,
\label{eq:GZMSSM}
\eeq
where $\Gamma_Z^{RSM}$ involves the contribution from a so-called
``Reference Standard Model''(RSM): namely, the Standard Model with a
Higgs mass set equal to the mass $m_{h^0}$ of the lightest
$CP$-even Higgs scalar of the MSSM  whereas
$\delta\Gamma_Z^{MSSM}$\,\cite{GJS1,GJS2}
constitutes the total quantum departure of the MSSM prediction
with respect to that RSM\footnote{For fully fledged analytical
formulae on $\delta\Gamma_Z^{MSSM}$ and an enlarged numerical analysis,
see the more comprehensive exposition of Ref.\cite{GJS3}.}.
Besides, $\delta\Gamma_Z^{MSSM}$ itself
splits up naturally into two parts, viz.
the extra two-doublet Higgs contribution $\delta\Gamma_Z^{H}$, in which
the single Higgs part included in $\Gamma_Z^{RSM}$ has been subtracted out
in order to avoid
double-counting, and the SUSY contribution, $\delta\Gamma_Z^{SUSY}$,
from the plethora of ``genuine'' ($R$-odd) supersymmetric particles:
\begin{equation}
\delta\Gamma_Z^{MSSM}=\delta\Gamma_Z^{H}+\delta\Gamma_Z^{SUSY}\,.
\label{eq:Tshift}
\end{equation}
Similarly for the ratio (\ref{eq:R}),
we may decompose the MSSM theoretical prediction as
follows:
\beq
R^{MSSM}=R^{RSM}+ \delta R^{MSSM}\,,
\label{eq:RhMSSM}
\eeq
where in an obvious meaning
\beq
\delta R^{MSSM}=\delta R^H+\delta R^{SUSY}=
R^{RSM}\,\left({\delta\Gamma_h^{MSSM}\over \Gamma_h^{RSM}}
-{\delta\Gamma_e^{MSSM}\over \Gamma_e^{RSM}}\right)\,.
\label{eq:dRhMSSM}
\eeq
Now, to compute the RSM contribution we shall adapt the convenient formula
of
Ref.\cite{Fit} for $R^{SM}$, in the following way:
\begin{eqnarray}
R^{RSM}&=&19.968\,(1-2.4\,10^{-4}\,\ln{m_{h^0}^2\over M_Z^2})\,
(1-2.5\,10^{-4}\,{m_t^2\over M_Z^2})\,
\left[\right.1+a_1\ {\alpha_s(M_Z)\over\pi}\nonumber\\
& &+a_2\,
\left({\alpha_s(M_Z)\over\pi}\right)^2
+a_3\,\left({\alpha_s(M_Z)\over\pi}\right)^3\left.\right]\,,
\label{eq:fit}
\end{eqnarray}
with
\beq
a_1=1.060\,,\ \ \ a_2=0.90-0.002\,(m_t/GeV-150)\,,\ \ \ a_3=15\,.
\label{eq:as}
\eeq
We have identified $R^{SM}$ with $R^{RSM}$ and straightforwardly
incorporated the
explicit dependence on the top quark mass and on
the (CP-even) Higgs mass. For the latter we have set $m_H=m_{h^0}$
and consistently related it with the mass of
the pseudoscalar, $m_{A^0}$, in the usual way predicted by the
MSSM\,\cite{Hunter}.

The very compact formula (\ref{eq:fit}) is a fit to the exact theoretical
calculation.
It summarizes all the complicated quantum effects from conventional
QCD up to 3-loop order and standard 1-loop
electroweak physics up to leading 2-loop order,
including the dominant $2$-loop strong-electroweak mixed effects,
and is accurate to within $0.0005$ in $\alpha_s(M_Z)$ over a range
$0.10 \stackrel{\scriptstyle <}{{ }_{\sim}}\alpha_s(M_Z)
 \stackrel{\scriptstyle <}{{ }_{\sim}}0.15$
for Higgs and top quark masses in the ranges
 $50\,GeV \stackrel{\scriptstyle <}{{ }_{\sim}} m_{h^0}
 \stackrel{\scriptstyle <}{{ }_{\sim}}1000\,GeV$ and
$100\,GeV \stackrel{\scriptstyle <}{{ }_{\sim}} m_t
\stackrel{\scriptstyle <}{{ }_{\sim}}200\,GeV$\,\cite{Manel}.

{}From eqs.(\ref{eq:RhMSSM})-(\ref{eq:as}) we numerically
determine $\alpha_s(M_Z)$ in the MSSM upon equating $R^{MSSM}$
\footnote{ In the analytic formulae for $\delta R^{MSSM}$\,\cite{GJS3},
we do not incorporate the 2-loop non-standard
Higgs effects on $\delta R^{H}$. They are expected to be
very small, in part because of the tight SUSY constraints in the Higgs
sector.}
to the experimental value\,\cite{Martinez1}
\beq
R^{\rm exp}=20.795\pm 0.040\,.
\eeq

Throughout our numerical analysis we shall survey the general MSSM
parameter
space using the $8$-tuple procedure devised in Ref.\cite{GJS2}:
\beq
(\tan\beta, m_{A^0}, M, \mu, m_{\tilde{\nu}}, m_{\tilde{u}_L},
 m_{\tilde{b}_L}, M_{LR})\,,
\label{eq:TUPLE}
\eeq
where it is understood that
the SUSY parameters in (\ref{eq:TUPLE})
will be picked from the typical intervals given in eq.(14) of
Ref.\cite{GJS2}.
For example, in Fig.1 we plot contour lines of constant $\alpha_s(M_Z)$
in a sufficiently large window of the $(\mu, M)$-parameter space,
for fixed $\tan\beta=50$ and $m_{A^0}=50\,GeV$, and for a given set of
sfermion
masses in the basic tuple (\ref{eq:TUPLE}). We see that the electroweak
SUSY effects are capable of significantly reducing the conventional
lineshape result (\ref{eq:alphahigh}) down to the low-energy
value (\ref{eq:alphalow}). Notice that there is a excluded
region of parameter space (see the shaded area in Fig.1a)
which comes about because we have subordinated
our analysis of $\alpha_s(M_Z)$ to the MSSM prediction of the total
$Z$-width,
$\Gamma_Z$\,\cite{GJS1}. The latter is experimentally bound to lie within
the
interval\,\cite{Glasgow}
\beq
\Gamma_Z^{\rm exp}=2.4974 \pm 0.0038\,GeV\,,
\label{eq:GZ}
\eeq
whereas the SM prediction, which we identify with $\Gamma_Z^{RSM}$,
reads\,\cite{Hollik}
\beq
\Gamma_Z^{\rm SM}=2.4922 \pm 0.0075\pm 0.0033\,GeV\,.
\label{eq:GZSM}
\eeq

Imposing the condition that $\delta\Gamma_Z^{MSSM}$  should
not exceed to $1\sigma$ the experimental value (\ref{eq:GZ}), with all
errors
(experimental and theoretical) added in quadrature, we find a forbidden
area in the $(\mu,M)$-plane of Fig.1a. Similarly, in Fig.1b we consider
the contour lines corresponding to a larger
$\tan\beta$ and larger sfermion masses. In this case the
shaded (excluded) area is much smaller, since
we have selected a set of parameters yielding small universal
contributions to $\Gamma_Z$, while the vertex contributions--specially from
the neutral Higgs sector--remain fairly large.

Let us study the dependence on the various parameters.
For instance, for a given set (\ref{eq:TUPLE}),
the largest possible values of $\tan\beta$ render the smallest
lineshape determinations of
$\alpha_s(M_Z)$. However, one should not exceed the upper bound of
$\tan\beta$, which roughly lies at $\tan\beta=70$.
A significant decrease of $\alpha_s(M_Z)$
can also be obtained for very large (decoupled)  $m_{A^0}$ and
very small $\tan\beta<0.7$. In both (extreme) regimes of $\tan\beta$
one is bordering the limit of perturbation theory
for the supersymmetric Yukawa couplings of charginos and neutralinos with
the stop-sbottom sector\footnote{The detailed interaction Lagrangian is
given
e.g. in eqs.(18)-(19) of Ref.\cite{GJSH}.}:
\begin{equation}
h_t={g\,m_t\over \sqrt{2}\,M_W\,\sin{\beta}}\;\;\;\;\;,
\;\;\;\;\; h_b={g\,m_b\over \sqrt{2}\,M_W\,\cos{\beta}}\,.
\label{eq:Yukawas1}
\end{equation}

In spite of the $\Gamma_Z$-constraint,
we see from Figs.1a and 1b that a wide region of parameter space is
phenomenologically accessible. Part of this region, however, is not so
interesting. In fact,
although large values of $|\mu|$ are permitted by the
$\Gamma_Z$-constraint,
the most relevant domain of the $(\mu,M)$-plane
is the one characterized by the smallest possible values of
$|\mu|$ compatible with the phenomenological bounds. The reason is that,
in this region, charginos have a
large higgsino component and are relatively light, viz. of ${\cal
O}(50)\,GeV$,
and in these conditions one can simultaneously improve the MSSM prediction
of $R_b$ and $R_c$\,\cite{GJS1}-\cite{GS}.

Another important feature of our analysis is the demand for a
relatively light ${\cal O}(50)\,GeV$ CP-odd scalar. At high $\tan\beta$,
it goes accompanied (in the MSSM) with a CP-even Higgs of about the same
mass.
Even though several lower mass limits on two-doublet Higgs sectors exist in
the literature,
the region $m_{A^0} \stackrel{\scriptstyle >}{{ }_{\sim}}40\,GeV$
has not yet been convincingly excluded by LEP
for $\tan\beta>>1$\,\cite{Yellow}.
In fact, in contradistintion to the usual analyses of mass limits on the SM
Higgs,
the couplings of the MSSM Higgs sector\,\cite{Hunter} are such
that the corresponding standard Bjorken decay
$Z\rightarrow h^0\,l^+ l^-$\,\cite{Bjorken}
for the lightest CP-even Higgs scalar in the MSSM
cannot be used to set any stringent lower limit on its mass
in the region of large $\tan\beta$.
On the other hand, the complementary decay
$Z\rightarrow h^0\, A^0$,
whose branching ratio increases dramatically at large $\tan\beta$,
is rendered inefficient for
$m_{A^0}\stackrel{\scriptstyle >}{{ }_{\sim}}
(0.3-0.4)\,M_Z$\,\cite{Yellow}.
As a result the whole range
$m_{A^0}\stackrel{\scriptstyle >}{{ }_{\sim}} 40\,GeV$ remains
open at high $\tan\beta$
and therefore we are free to exploit the window around the lower band
of permitted values of $m_{A^0}$ where the interesting
effects take place\,\cite{GJS2}.

As for the demand for light (higgsino-like) chargino-neutralinos
of $50-60\,GeV$\,\cite{Tenyears}, it is
worthwhile remembering that they are imperative not only
because they may contribute sizeable non-oblique corrections
to some partial $Z$-widths,
but also because they produce large, and {\it negative},
oblique effects on all partial
 $Z$-widths. Since the oblique effects essentially
cancel in the ratios (\ref{eq:Rb}),(\ref{eq:Rc}), the latters are dominated
by just the non-oblique corrections. The upshot of this game is that
one may generate a pattern of quantum corrections
in the MSSM characterized by non-negligible
effects on $R_b$ and $R_c$ while at the same time leaving a small enough
net correction on the total $Z$-width
 (Cf. Fig.1 of Ref.\cite{GJS1})
that respects the tight $\Gamma_Z$-constraint.

In Figs.2a-2b we plot the evolution of our corrections as a function of the
pseudoscalar mass, $m_{A^0}$, for three large values of $\tan\beta$
in two antipodal points of the $(\mu,M)$-plane.
It becomes clear that a light pseudoscalar mass
$m_{A^0}\stackrel{\scriptstyle <}{{ }_{\sim}}70\,GeV$
is favoured by a possible solution to the
 ``$\alpha_s(M_Z)$ crisis'' at high $\tan\beta$.
We see that, for $\tan\beta>50$, we are able to approach the
desired $\alpha_s(M_Z)=0.11$ regime on the basis of
pure electroweak corrections alone.
Finally, in Figs.3 and 4 we test the sensitivity of $\alpha_s(M_Z)$ on
the sfermion spectrum and in particular on the
mixing parameter of the stop mass matrix. Specifically, in Fig.3a we
exhibite the (mild) sensitivity on $m_{\tilde{u}}$ (assumed to be
degenerate with
$m_{\tilde{c}}$), and in Fig.3b we plot the (also light) dependence on the
slepton masses through the sneutrino mass parameter $m_{\tilde{\nu}}$.
We deal separately in Fig.4 with the
stop-sbottom case, due to the
likely presence of mixing in the stop sector.
{}From Fig.4a we verify
that $\alpha_s(M_Z)$ noticeably decreases with a decreasing sbottom mass.
This was expected from the relatively
large, and {\it positive}, supersymmetric contribution
to the $b\bar{b}$ decay mode which, at very high $\tan\beta$,
is dominated by the total yield from the pseudoscalar plus
sbottom-neutralino vertex diagrams. Such a contribution overwhelms that of
the stop-chargino vertex and renders the influence from the
stop mass rather irrelevant.
Indeed, for fixed sbottom mass, Fig.4b confirms that
 $\alpha_s(M_Z)$ is insensitive to  $M_{LR}$, and so to the stop masses,
thus proving our contention that a light stop
is not essential in our analysis. This is in stark contrast to the
low $\tan\beta$ regime, where the stop-chargino vertex
takes off and $\alpha_s(M_Z)$
becomes very sensitive to $M_{LR}$.

Some discussion is worth doing concerning the SUSY-QCD corrections as
compared to the SUSY-electroweak corrections.
For the formers to produce the desired quantum effect
one needs\,\cite{Djouadi} very light values for the gluino
masses, if squark masses are of the order of ${\cal O}(100)\,GeV$.
This possibility, even though
not fully excluded as long as gluinos of ${\cal O}(1)\,GeV$ remain
phenomenologically viable\,\cite{Clavelli}, is almost ruled out.
Alternatively,
our calculation proves that the relevant effect could also drop from pure
electroweak SUSY quantum physics. Remarkably enough, these electroweak
supersymmetric corrections
have two distinctive virtues, to wit: they are not only potentially
larger than the strong supersymmetric corrections,
but they even show the suitable alternation of signs so as to
simultaneously
adjust the $R_b$ and  $R_c$ theoretical predictions.
This feature is not possible
for the SUSY-QCD corrections\,\cite{Djouadi}, since they are always
positive for
all light-quark channels, in particular for the $c\bar{c}$ channel, and are
either large and negative, or small and positive, for the $b\bar{b}$
channel. Thus they could contribute significantly both to $R_b$ and $R_c$
just in the wrong
direction. This is specially so when $R_b$ receives
large negative corrections--a circumstance which is tied to the
hypothetical
existence of light sbottoms. In this case, strong SUSY-QCD effects
from sbottoms would dangerously
interfere with the positive contributions from the SUSY-electroweak
sector.

To prevent these negative corrections from occurring,
we have assumed that the mixing parameter in the sbottom sector,
$m_b(A_b-\mu\tan\beta)$, is small enough
so that light sbottoms of order $45\,GeV$ are forbidden
\footnote{Alternatively, gluinos could be very heavy and the SUSY-QCD
contributions would totally decouple.}. In contrast,
in the SUSY-QCD calculation of Ref.\cite{Djouadi}, light sbottoms are
allowed and go associated with large values of $\tan\beta$, the reason
being that
the trilinear soft SUSY-breaking parameters, $A_{b,t}$, have been
arbitrarily set to zero.
In general, however, the trilinear terms do not vanish and one
may have large
sbottom masses even for large $\tan\beta$. In this case the negative
SUSY-QCD contributions on the $b\bar{b}$ mode are safely restrained and
they
do not spoil the positive SUSY-electroweak effects presented here.

In summary, we have demonstrated the existence of regions of
the general MSSM parameter space
where electroweak supersymmetric quantum physics could
reveal itself as the ``new physis'' called for to simultaneously
resolve the  triple ``$R_b-R_c-\alpha_s(M_Z)$ crisis''.
These regions are characterized either by
very large or--less likely--by very small values of $\tan\beta$ and also,
respectively,
by a light CP-even and a light CP-odd Higgs of the same mass,
or by a very large (effectively decoupled) CP-odd Higgs.
In both cases a light chargino-neutralino is mandatory. However,
whereas in the low $\tan\beta$ region
a light stop of ${\cal O}(50) GeV$ is also indispensable,
in the high $\tan\beta$ region it is not needed at all; instead, an
intermediately heavy sbottom of a few hundred $GeV$
suffices to round off a comfortable solution to the triple crisis in the
MSSM.
As an extra bonus to be added up to our approach, we remark that
the solution space just described does automatically preserve
the successful SM prediction of the observables $M_W$,
$sin^2{\theta}_{\rm eff}^l$ and $A_{FB}^{0\,b}$\,\cite{Martinez1}.
In other words, our RSM prediction for these observables is essentially
the same as in the SM. The reason is simple: it stems from the fact that
these observables are dominated by universal corrections and hence are
weakly sensitive to the Higgs sector of the MSSM.
Furthermore, $M_W$, $sin^2{\theta}_{\rm eff}^l$ and $A_{FB}^{0\,b}$
are fairly blind, up to small
L-R mixing effects, to the presence of RH stops and sbottoms which, due to
the higgsino-like nature of the lightest chargino-neutralino,
are the only ones involved in the  relevant $Z$ vertices.

Despite it is not strictly necessary from the phenomenological point of
view,
it remains to see--if only as a standard
theoretical prejudice--whether our favourite parameter region can be
successfully
implemented in some (non-minimal) SUSY-GUT scenario. For example, in the
string type unified scenario of Ref.\cite{Roszkowski} one finds that
in order that the grand
unified prediction of $\alpha_s(M_Z)$ be consistent as much as possible
with
the low-energy value (\ref{eq:alphalow}), gluinos and squarks
should be relatively light, viz. of ${\cal O}(100)\,GeV$, and the wino
soft-mass parameter, $M$, should be much larger than the gluino mass.
Notice that this feature is not incompatible with a chargino being mostly
higgsino, as currently required in the MSSM literature on $R_b$.
Furthermore, from the contour lines of $\alpha_s(M_Z)$ in the
$(\mu, M)$-plane of Fig.1a we see that,
for a given value of $\mu$,
it is precisely in the region of large $M$ where $\alpha_s(M_Z)$
approaches the closest to the low-energy value, eq.(\ref{eq:alphalow}),
while
at the same time preserves the experimental bound (\ref{eq:GZ})
on $\Gamma_Z^{\rm exp}$.
As an optimized
example, we take $\mu=-50\,GeV,\, M=350\,GeV$
and the rest of the parameters as in Fig.1b, and we find
 $\alpha_s(M_Z)\simeq 0.111$, thus a value which is in excellent agreement
with the deep inelastic scattering result (\ref{eq:alphalow})
and being also compatible with the rest of the $Z$-observables.
Still, it is true that our analysis favours the higgsino together with the
lightest Higgs to be in the ${\cal O}(50)\,GeV$ range, a fact which
could be at variance with the GUT spectrum derived in
Ref.\cite{Roszkowski}.
Clearly, more work is needed in the arena of supersymmetric grand unified
theories before reaching a conclusive solution to the ``$\alpha_z(M_Z)$
crisis''
that is compatible with the experimental status of the various $Z$
observables.
Be as it may, if this status remains unchanged in the next generation of
experiments, a SUSY-GUT solution within the MSSM, if it exists at all,
is bound to meet the general conditions reported in the present study.

{\bf Added Note}

After completing our work, we became aware of a work by
Chankowski and Pokorski\,\cite{CP} which projects the
same preferred region of the MSSM parameter space as in our case, and a
work
by Kane, Stuart and Wells\,\cite{KSW} where a fit analysis of the data is
made
on the basis of {\it assuming} (not computing!) the
low energy value (\ref{eq:alphalow}) of $\alpha_s(M_Z)$
in the MSSM. For comparison purposes, it
may be useful to note that the latter work by Kane et al. does not
incorporate the
effect from the Higgs pseudoscalar contributions  and as
a consequence their approach is not adequately sensitive to the high
$\tan\beta$
region. This leads them to overemphasize the-- potentially problematic--
existence of a too light
stop, a possibility which, although sufficient, is not necessary to fit the
electroweak data. In fact, as we have shown in the present and previous
analyses\,\cite{GJS1}-\cite{GS}, a complete treatment of the quantum
effects
from the MSSM with a Higgs sector containing a light pseudoscalar
(with or without a light stop) may
be crucial to account for the electroweak data on the $Z$,
as also confirmed by the aforementioned work by Chankowski and Pokorski.

\newpage
{\bf Acknowledgements}

One of us (JS) is indebted to M. Mart\'\i nez for providing details on
the fit analysis of Ref.\cite{Fit} which we have used in this paper.
He is also obliged to M. Shifman for
reading the manuscript and for sharing his insight on the QCD implications
of a high $\alpha_s(M_Z)$. Conversations with P. Chankowski are also
gratefully acknowledged.
This work has  been partially supported by CICYT
under project No. AEN93-0474. DG has also been financed by a grant of
the Comissionat per a Universitats i Recerca, Generalitat de Catalunya.


\newpage
\vspace{1cm}
\begin{center}
\begin{Large}
{\bf Figure Captions}
\end{Large}
\end{center}
\begin{itemize}
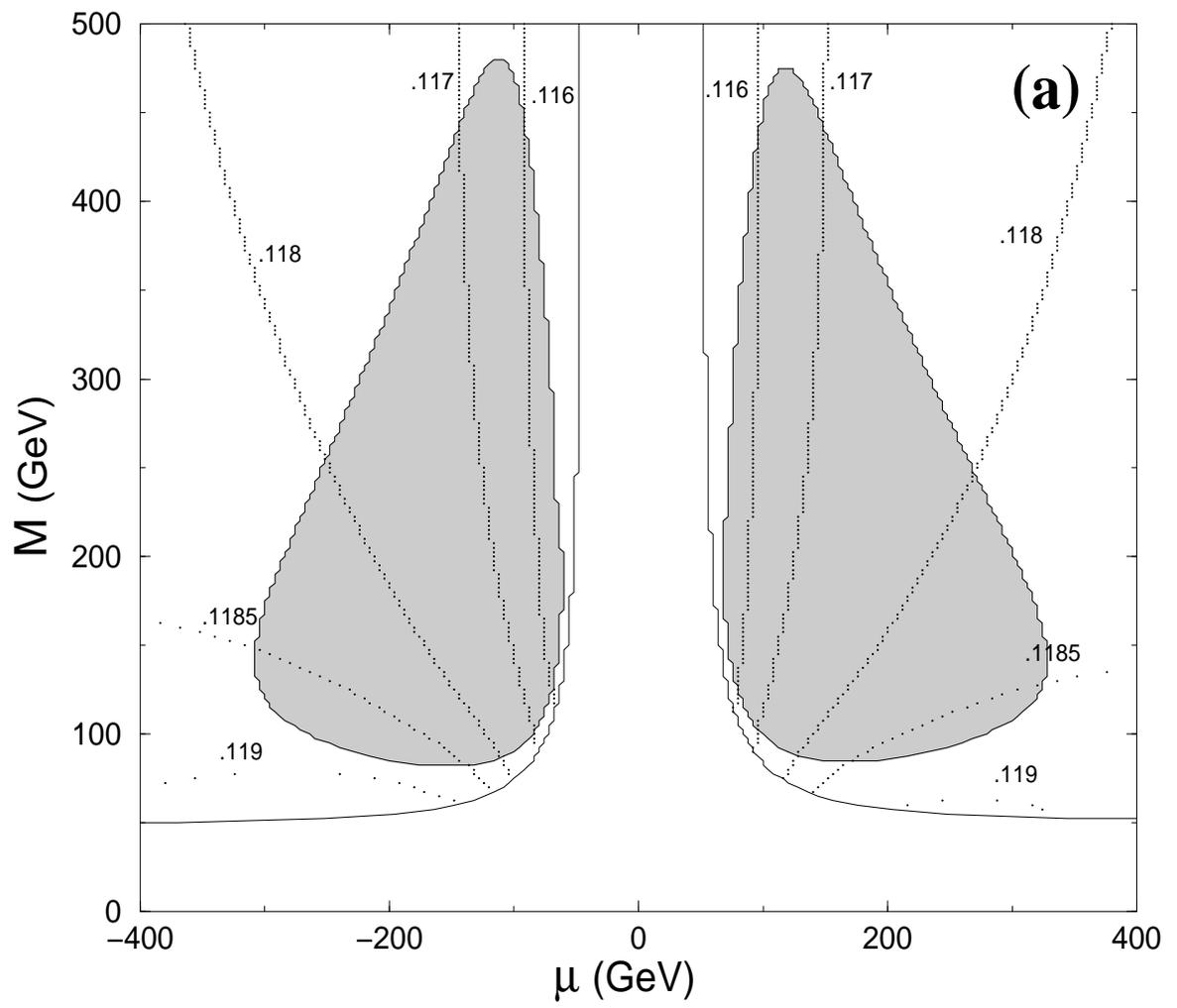
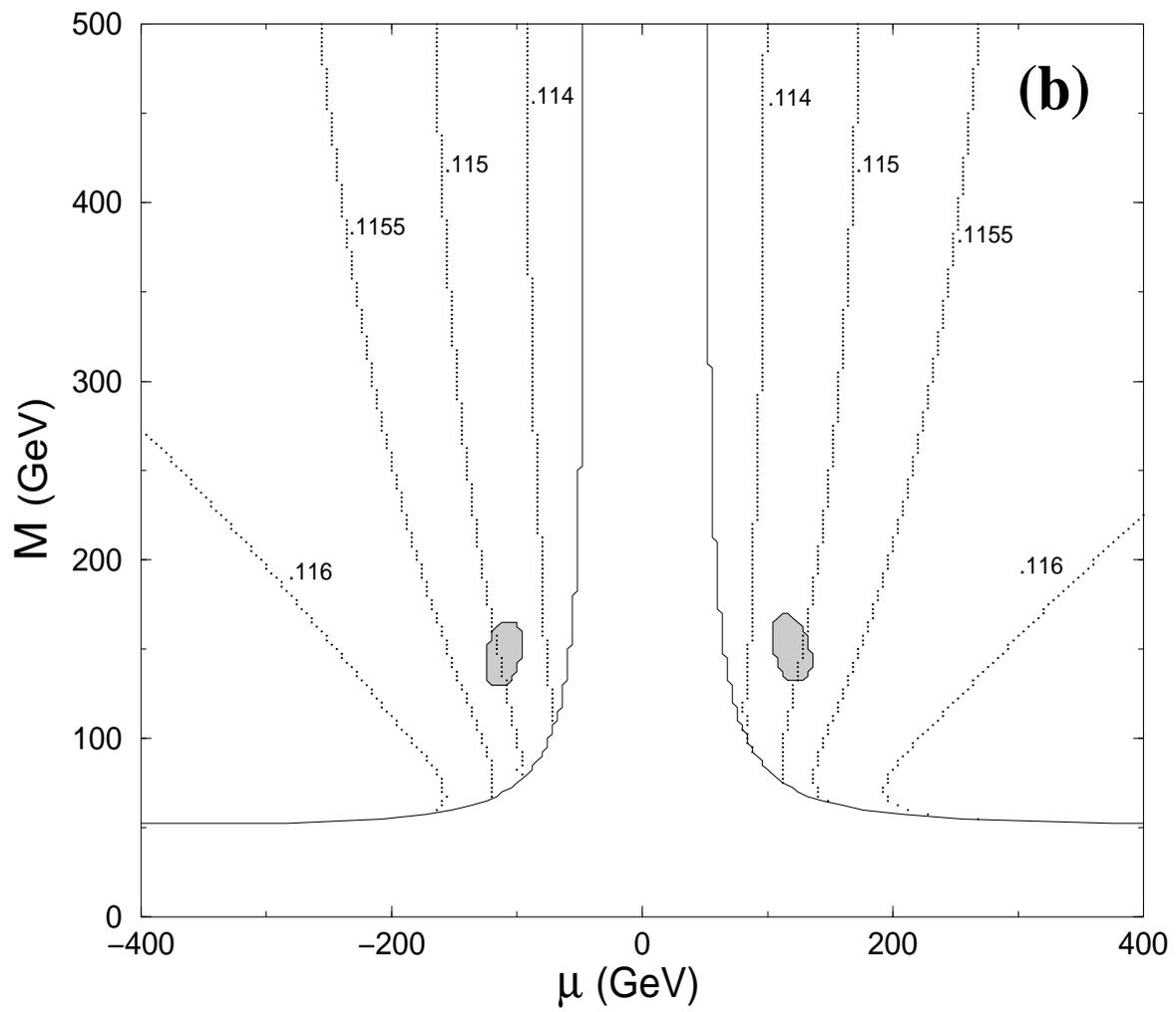
\item{\bf Fig.1} (a) Contour lines for the lineshape-determined
$\alpha_s(M_Z)$ in the
higgsino-gaugino $(\mu, M)$-plane of the MSSM.
The sfermion spectrum is obtained from the basic parameter set
(\ref{eq:TUPLE}) following the procedure of Ref.\cite{GJS2} with
$\tan\beta=50$, $m_{A^0}=50\,GeV$,
$m_{\tilde{\nu}_l}=m_{\tilde{u}}= m_{\tilde{b}}=200\,GeV$ and $M_{LR}=0$.
The top quark mass is fixed at $m_t=175\,GeV$. The blank region is
excluded by the chargino-neutralino mass bounds\,\cite{GJS1}, and
the shaded area is excluded by the $\Gamma_Z$-constraint:
$\delta\Gamma_Z^{MSSM}<9\,MeV$;
(b) As in case (a), but for $\tan\beta=60$ and
$m_{\tilde{\nu}_l}=m_{\tilde{u}}= m_{\tilde{b}}=300\,GeV$.

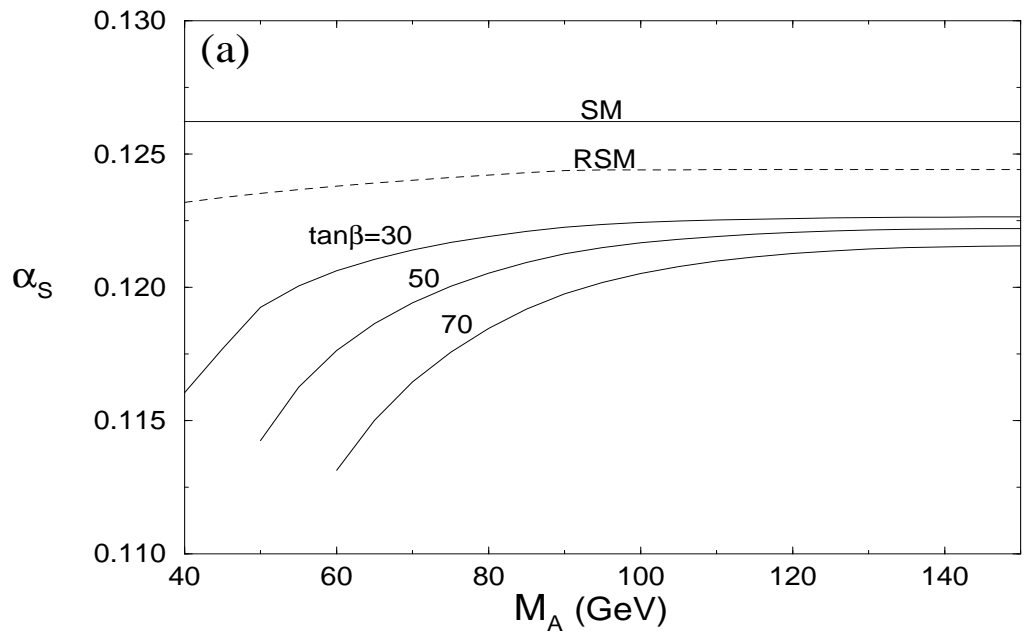
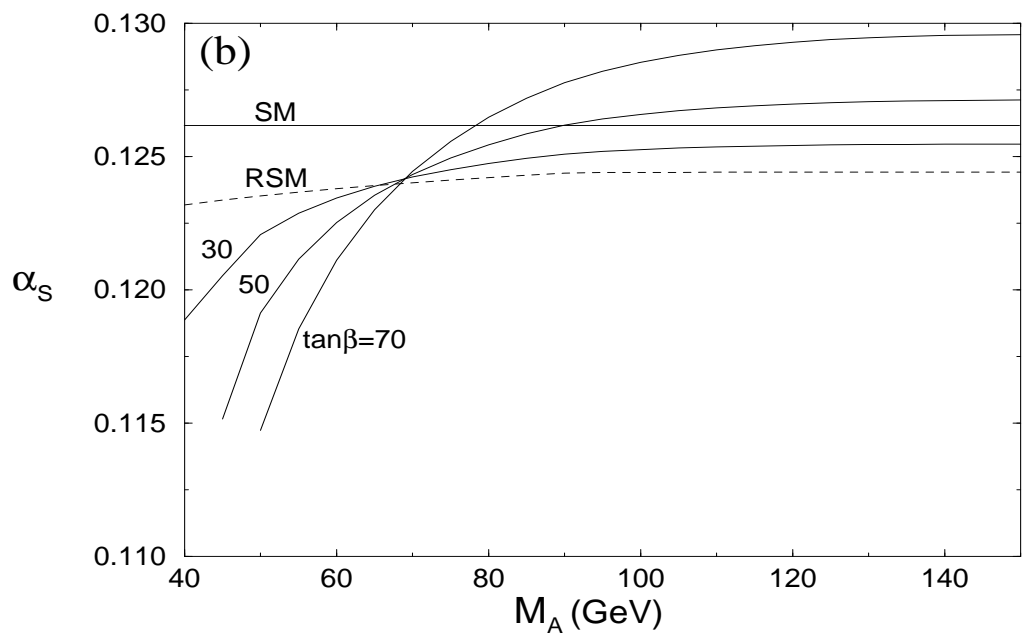
\item{\bf Fig.2} (a) $\alpha_s(M_Z)$ as a function of the
pseudoscalar mass $m_{A^0}$ for various $\tan\beta$,
$(\mu,M)=(-60,350)\,GeV$
and the same sfermion spectrum as in Fig.1a. The curves are cut off
from below by the $\Gamma_Z$-constraint; (b) As in case (a), but
for $(\mu,M)=(-350,60)\,GeV$. The dashed curve corresponds to the
RSM contribution alone, and the straight line to the SM with a fixed Higgs
mass of $m_H=300\,GeV$.

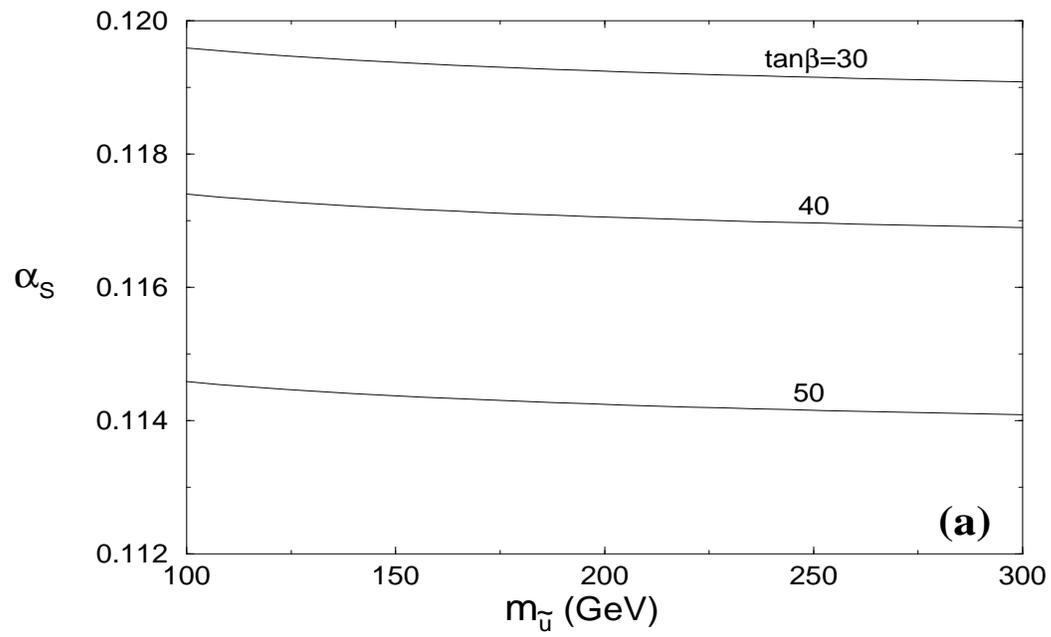
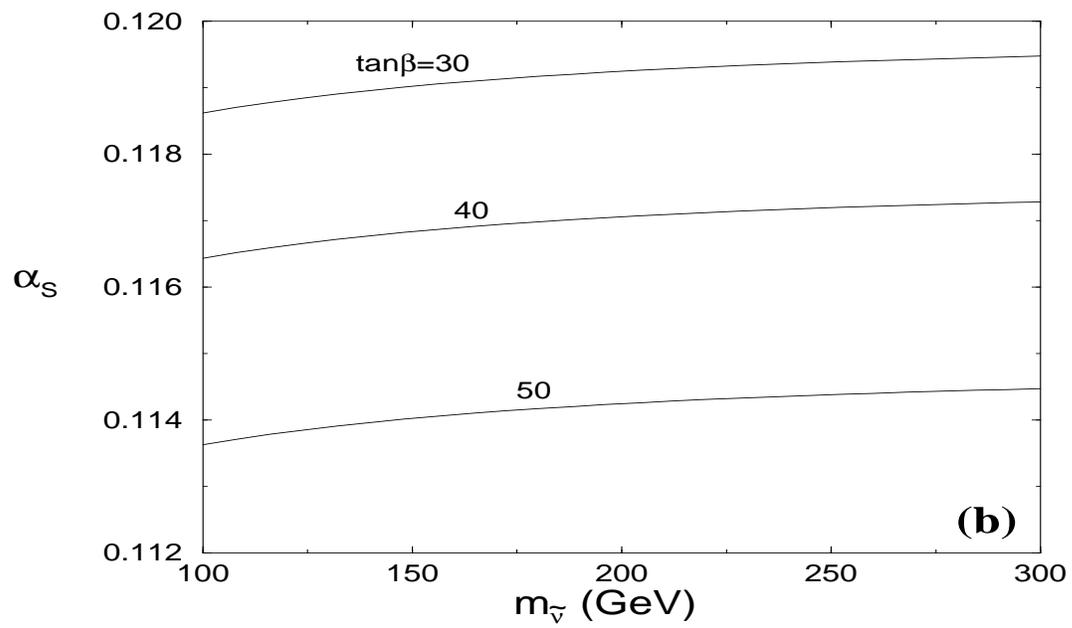
\item{\bf Fig.3} (a)  $\alpha_s(M_Z)$ as a function of the squark masses
$m_{\tilde{u}}=m_{\tilde{c}}$ for various $\tan\beta$. The remaining
sfermion masses are as in Fig.1a, and $(\mu,M)$ fixed as in Fig.2a;
(b) $\alpha_s(M_Z)$ as a function of the sneutrino
masses $m_{\tilde{\nu}}$ and the rest of parameters as in case (a).

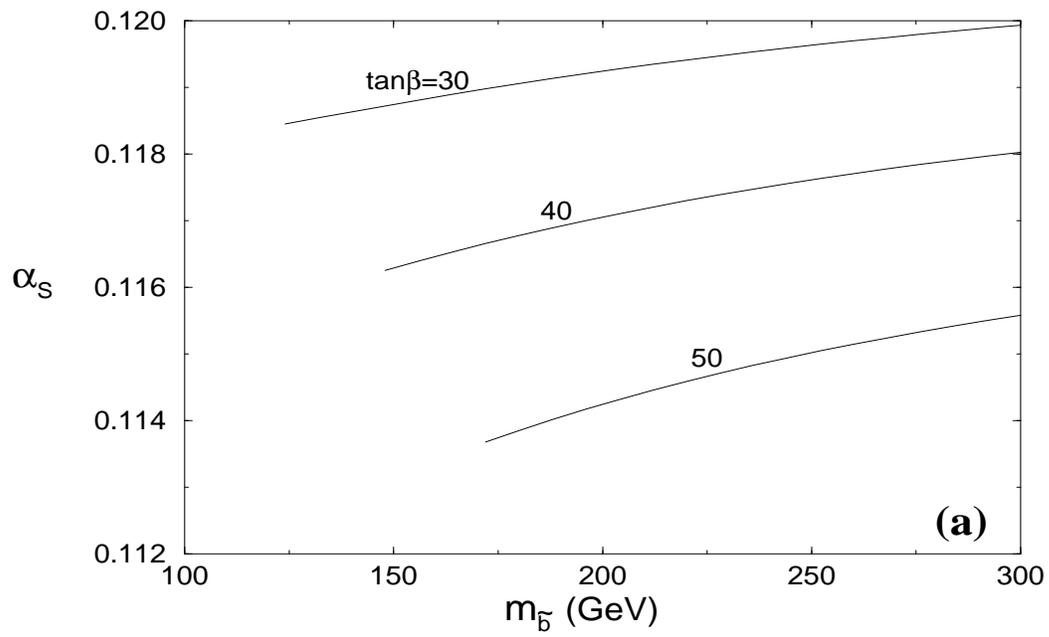
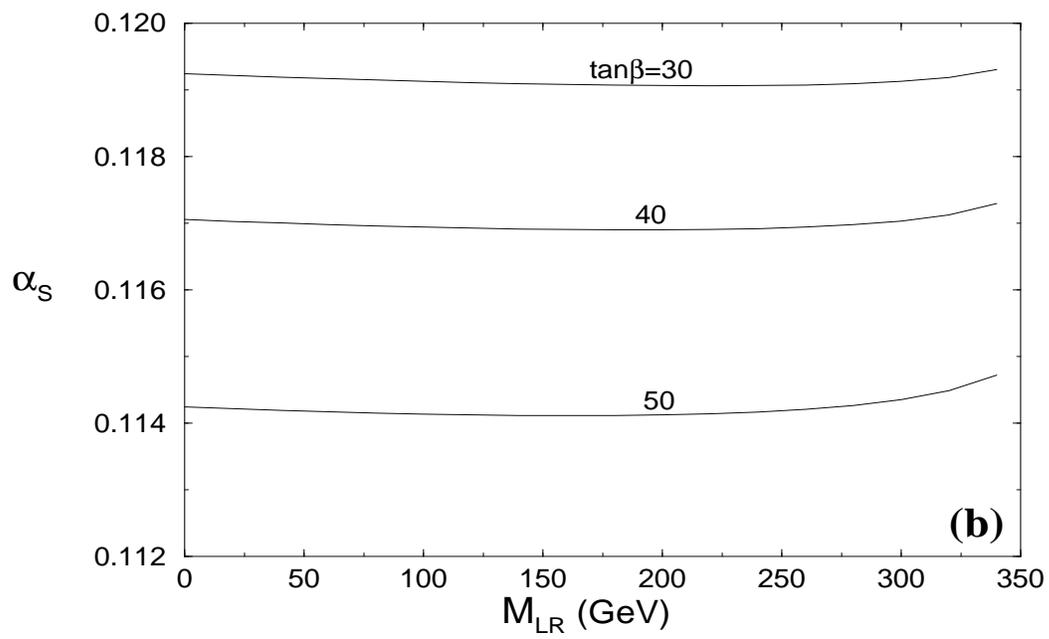
\item{\bf Fig.4}  (a) Evolution of $\alpha_s(M_Z)$ in terms of
$m_{\tilde{b}_L}=m_{\tilde{b}_R}\equiv m_{\tilde{b}}$,
for $(\mu,M)$ as in Fig.2a and the rest
of the parameters as in Fig.1a;
(b) Dependence of $\alpha_s(M_Z)$ on
the stop mixing parameter $M_{LR}$ for the same choice of the
remaining parameters as in case (a).

\end{itemize}

\end{document}